\documentclass[journal]{IEEEtran}

\usepackage{amsmath,amssymb}
\usepackage{multicol}
\usepackage{graphicx}
\usepackage{caption}
\usepackage{color}

\usepackage{xcolor}

\title{Optimization of Long-Haul C+L+S Systems by means of a Closed Form EGN Model}
\author{Y. Jiang, J. Sarkis, A. Nespola, F. Forghieri, S. Piciaccia, A. Tanzi, 
M. Ranjbar Zefreh, P. Poggiolini
\thanks{Y. Jiang, J. Sarkis and P. Poggiolini are with Politecnico di Torino, Torino, Italy. A. Nespola is with Links Foundation, Torino, Italy. F.~Forghieri, S.~Piciaccia, A. Tanzi, and M. Ranjbar Zefreh are with CISCO Photonics, Vimercate (MB), Italy.}

\thanks{This work was partially supported by: Cisco Systems through the ITROCS research contract; the PhotoNext Center of Politecnico di Torino; the European Union under the Italian National Recovery and Resilience Plan (NRRP) of NextGenerationEU, partnership on “Telecommunications of the Future” (PE00000001 - program “RESTART”).}}
\date{February 2024}

\begin{document}

\maketitle
\begin{abstract}
We investigate C+L+S long-haul systems using a closed-form GN/EGN non-linearity model. We perform accurate launch power and Raman pump optimization. We show a potential 4x throughput increase over legacy C-band systems in 1000~km links, using moderate S-only Raman amplification. We simultaneously achieve extra-flat GSNR, within $\pm$0.5~dB across the whole C+L+S spectrum.
\end{abstract}
    

\section{Introduction}
Many technologies are currently competing in the quest for increasing the throughput of optical links. They can be broadly classified as either “space-division-multiplexing” (SDM) or “multi-band” (MB). All SDM and some MB technologies require that new cables be deployed. A notable exception is MB over existing standard single-mode fiber (SMF) cables, a potentially attractive alternative for carriers who want to exploit existing cables to their ultimate potential. MB over SMF consists of extending the transmission bandwidth beyond the C band. The first step, the extension to L-band, is already commercially available. Research is now focusing on other bands, primarily S and O but also E and U. In the context of long-haul systems, which this paper focuses on, it is mostly the S-band that is being considered, because higher frequency bands such as E and O suffer from more serious propagation impairments, while the U-band appears problematic due to bend loss and non-mature amplification solutions.

The nominal bandwidth of the S-band is quite large, almost 10~THz. Current efforts aim at exploiting the 5-6 THz adjacent to the C-band, about 196.5 to 202.5 THz, because propagation conditions are more favorable than at higher frequencies, and amplification is available as Thulium-doped fiber amplifiers (TDFAs). However, it is not inconceivable that in the future the upper limit can be pushed further. 
Many research experiments of C+L+S transmission have already been successfully carried out. For instance \cite{2023_ECOC_Frignac} where 18 THz (6 THz each for C, L, S) over 2x60km were transmitted. Also \cite{2023_ECOC_Hout} where, remarkably, 12,345 km were reached, using about 3 THz of S-band (plus C and L band with 6 THz each) over special low-loss 4-core MCF and Raman amplification with 8 pumps. Another  C+L+S example is \cite{2023_JLT_Escobar}, with 200~Tb/s over 2x100 km PSCF.

However, to achieve commercial attractiveness in conventional terrestrial long-haul, C+L+S systems must conceivably meet certain key goals: (a) C+L+S must bring about a very substantial throughput increase, such as 4x or more vs.~the still ubiquitous legacy C-band systems (4.4 to 4.8-THz bandwidth); (b) the operating conditions in the three bands should be rather uniform (similar GSNRs); (c) if used, Raman amplification should need a small number of limited power pumps.

To pursue these goals, joint optimization needs to be carried out of key system parameters, such as WDM launch power spectra and Raman pump frequencies and powers. This requires fast and accurate physical layer models, capable of accounting for the broadband-dependence of all fiber and system parameters, together with Inter-channel Raman Scattering (ISRS) and Raman amplification. Closed-Form Models (CFMs) have been developed for this purpose. Mainly two groups, one at UCL, and one at PoliTo (in collaboration with CISCO), have independently obtained CFMs based on approximations of the GN/EGN models, with similar foundations but with differences in features and final analytical form. For the UCL CFM see \cite{2023_JLT_Buglia}, \cite{2023_JLT_Buglia2}, for the CISCO-PoliTo CFM see \cite{2020_arXiv_RanjbarZefreh}, \cite{2022_ECOC_Poggiolini}, \cite{2024_arXiv_Jiang}. Extensive experimental validations of the CISCO-PoliTo CFM (henceforth just “CFM”) were presented at ECOC 2023 and 2024 \cite{2023_ECOC_Jiang, 2024_ECOC_Jiang}.

In this paper, we focus on a long-haul 1000~km SMF system, using about 18 THz for C+L+S transmission (about 6 THz per band), similar to \cite{2023_ECOC_Frignac}. Note that 6 THz is an extended bandwidth for C and L, sometimes called `super-C' and `super-L'. However, for brevity, we will drop the `super' qualifier. We carry out multi-parameter optimization to achieve the goals (a)-(c) listed above. We look at transmission with and without Raman. We optimize WDM launch power spectrum and Raman pump power and frequency, aiming at maximizing throughput, also subject to GSNR uniformity. We show that a greater than 4x throughput increase vs.~standard C-band systems appears achievable in 1000km links, within a ±0.5dB GSNR uniformity across all 18 THz of spectrum, using only 3 Raman pumps and a total of less than 1~Watt of pump power.
A preliminary version of this investigation was presented at OFC 2024 \cite{2024_OFC_Jiang}. Here we greatly expand on the results, especially on Raman-supported systems, on the details and discussion.

\begin{figure*}
    \centering
    \includegraphics[width=0.75\linewidth]{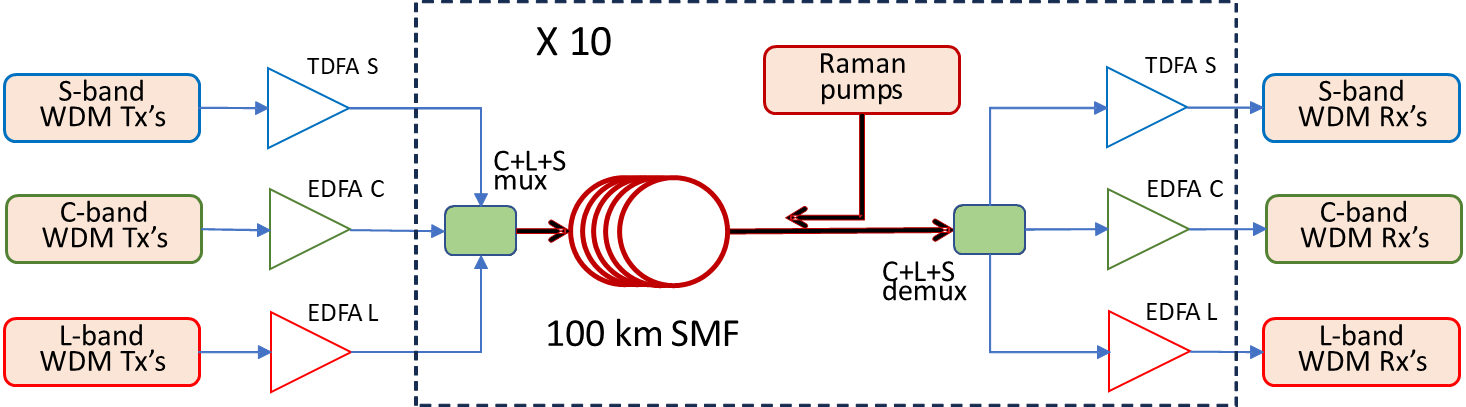}
    \caption{Schematic of the 10-span C+L+S system being studied. Fiber parameters are slightly different among spans (see text).}
    \label{fig:schematic}
            \vspace{-0.3cm}
\end{figure*}

\section{System Description and Results}
\begin{figure}
    \centering
    \includegraphics[width=0.8\linewidth]{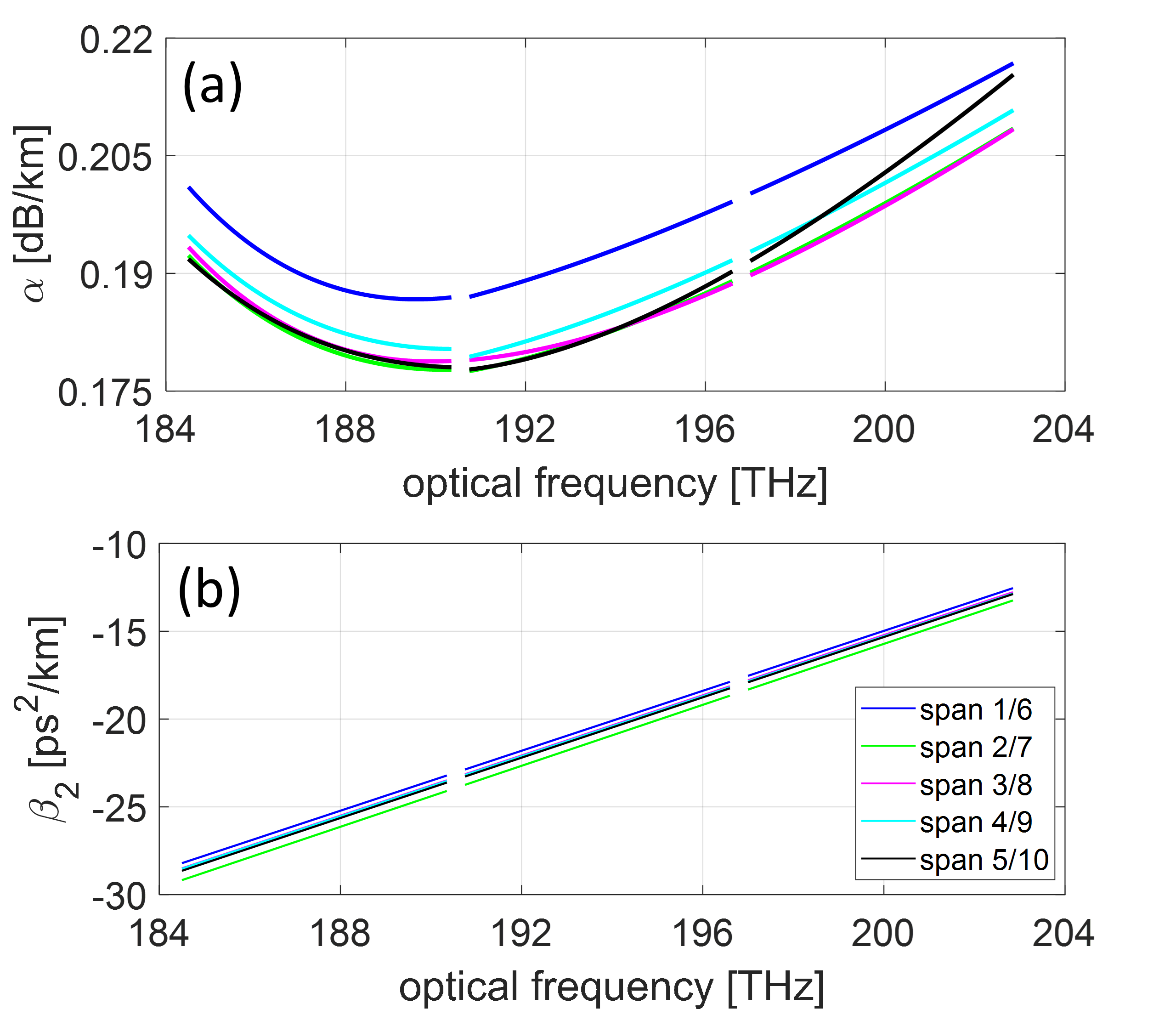}
    \caption{Loss and dispersion, measured in C and L band and extrapolated to S band.}
    \label{fig:loss_and_dispersion}
        \vspace{-1cm}
\end{figure}

\begin{figure}
    \centering
    \includegraphics[width=0.9\linewidth]{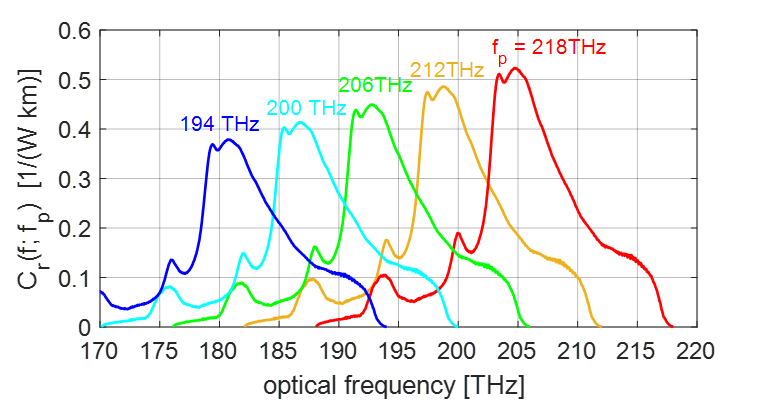}
    \caption{Raman gain spectrum $C_{\rm R}$ measured for a 206.5~THz pump and then shifted and scaled according to pump frequency \cite{2003_JLT_Rottwitt}. }
    \label{fig:Raman}
    \vspace{-1cm}
\end{figure}

\begin{figure}
    \centering
    \includegraphics[width=0.8\linewidth]{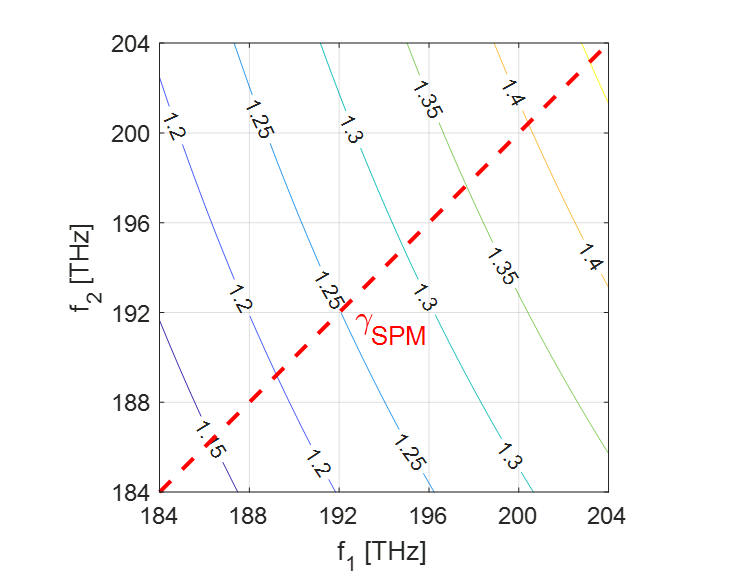}
    \caption{Contour plot of the cross-channel interference (XCI) non-linearity coefficient $\gamma\left(f_1,f_2\right)$ 1/(W$\cdot$km). The single-channel interference (SCI) value is found for $f_{1}\!=\!f_{2}$ (dashed red line) 
    \cite{2022_ECOC_Poggiolini}, \cite{2010_OE_Santagiustina}. }
    \label{fig:gamma}
\end{figure}

The schematic is shown in Fig.\ref{fig:schematic}. It comprises 10 spans. The first five spans were characterized for the experimental set-up used for the CFM validation in \cite{2023_ECOC_Jiang}. Loss and dispersion were measured in C and L band and then extrapolated to the S-band (Fig.\ref{fig:loss_and_dispersion}) using well-known formulas \cite{1986_JLT_Walker}. The Raman gain spectrum $C_{\rm R} (f,f_p )$ was experimentally characterized using a pump at $f_p$=206.5 THz (Fig.\ref{fig:Raman}). It was shifted and scaled as a function of $f$ and at $f_p$ according to \cite{2003_JLT_Rottwitt}.  Fig.\ref{fig:gamma} shows the contour plot of the fiber non-linearity coefficient $\gamma\left(f_1,f_2\right)$, according to the formulas reported in \cite{2022_ECOC_Poggiolini}, \cite{2010_OE_Santagiustina}. The frequency $f_1$ is that of the channel-under-test (CUT) whereas $f_2$ is the frequency of the channel creating cross-channel interference (XCI) on the CUT. If $f_1\!=\!f_2$ the value of $\gamma$ for single-channel interference (SCI) is obtained (red dashed curve).  To evaluate both $\gamma\left(f_1,f_2\right)$ and $C_{\rm R} (f,f_p )$, the mode effective area is needed, and the expression in \cite{1977_BellTech_Marcuse} was used. 

\begin{figure}
    \centering
    \includegraphics[width=0.8\linewidth]{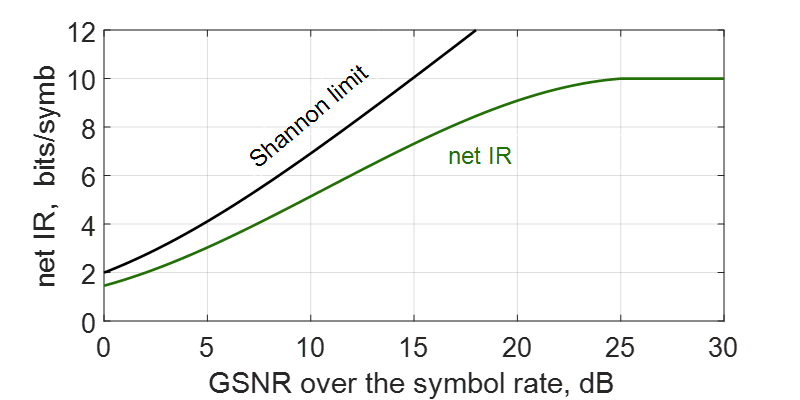}
    \caption{Transponder net user information rate vs.~GSNR.}
    \label{fig:IR}
                \vspace{-0.3cm}
\end{figure}

To obtain a 10-span set-up from the 5-span experiment \cite{2023_ECOC_Jiang}, the five spans were replicated, with identical parameters. Also, in \cite{2023_ECOC_Jiang} spans were on average 85km length. Here we analytically stretched them to 100 km so that the total span loss was about 22 dB per span at 190 THz, on average about 18.5 dB from fiber loss and the rest accounting for various lumped loss (connectors, band mux-demux, etc.). The exact WDM band boundaries were similar to \cite{2023_ECOC_Frignac}: L-band 184.50 to 190.35; C-band 190.75 to 196.60; S-band 197.00 to 202.85. Doped-Fiber-Amplifiers (DFAs) were assumed with 6dB noise-figure in L and S-band and 5dB in C-band, according to the experimental values \cite{2023_ECOC_Frignac}. The WDM signal consisted of 50 channels in each band, with 100 GBaud symbol rate, roll-off 0.1, and spacing 118.75 GHz. Modulation was assumed Gaussian-shaped. The \textit{net user information rate} of the transponders, accounting for FEC overhead, was assumed as shown in Fig.\ref{fig:IR}, representative of the latest generation of transponders coming to market in 2024.


Our goal was to perform launch power spectrum optimization, assumed to be the same into each span, to achieve throughput maximization. The objective function that we chose was:
\begin{equation}
    f_{\rm obj} = {\rm mean}\left( {\rm IR}_{\rm Rx}^{n} \right)
    \label{eq:objective_function}
\end{equation}      
where ${\rm IR}_{\rm Rx}^{n}$ is the information rate of the $n$-th channel, obtained from the GSNR through the curve of Fig.~\ref{fig:IR}.

\begin{figure*}[ht]
    \centering
    \includegraphics[width=0.95\linewidth]{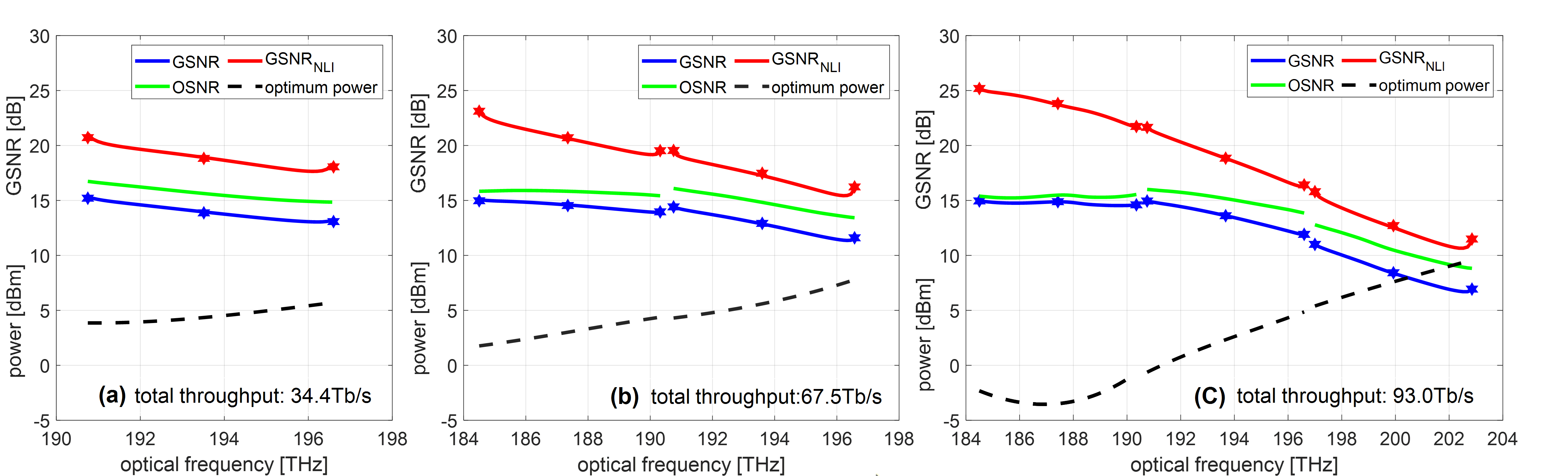}
    \caption{Optimum launch power and related $\rm OSNR$ (ASE only), $\rm GSNR_{NLI}$ (NLI only) and $\rm GSNR$ (ASE and NLI) for 10x100km SMF systems  using (a): C-band; (b): C+L; (c): C+L+S. Star markers: CFM accuracy-check by numerical integration of the multiband EGN model. }
    \label{fig:results}
\end{figure*}

In Fig.~\ref{fig:results}, we show the optimization results for C, C+L, and C+L+S systems. The optimized launch power spectrum is the black dashed line. The red solid curve represents the so-called `non-linear GSNR' (NLI only), the green solid curve represents OSNR (ASE only) and the blue solid line shows the overall GSNR, accounting for both ASE and NLI. The markers were obtained by numerically integrating the full EGN model. They were calculated to perform a CFM accuracy-check, consistently indicating excellent agreement between the CFM and the EGN model. Fig.~\ref{fig:results}(b) shows substantial signs of ISRS already in the C+L system, manifesting as optimum launch power going up towards the high frequencies. Nonetheless, throughput essentially doubles, from 34.4 to about 67.5 Tb/s, when going from C (Fig.~\ref{fig:results}(a)) to C+L. This aligns well with many known results, both simulative and experimental, that show that the impact of ISRS can be made almost negligible in optimized C+L systems.

When the S band is turned on, though, (Fig.~\ref{fig:results}(c)) throughput increases only by 25.5 Tb/s, achieving 93.0 Tb/s. This is not proportional to the amount of added bandwidth, which is therefore used less efficiently. The reason is the system GSNR dipping down in S-band, with the high-frequency S-band channels operating 8~dB below the L-band GSNR. In addition, the peak-to-peak variation in optimum launch power across bands is extremely large, about 13~dB. This may pose practical problems, especially for the high-frequency S-band channels which require 10~dBm launch power.

\begin{figure}
    \centering
    \includegraphics[width=0.88\linewidth]{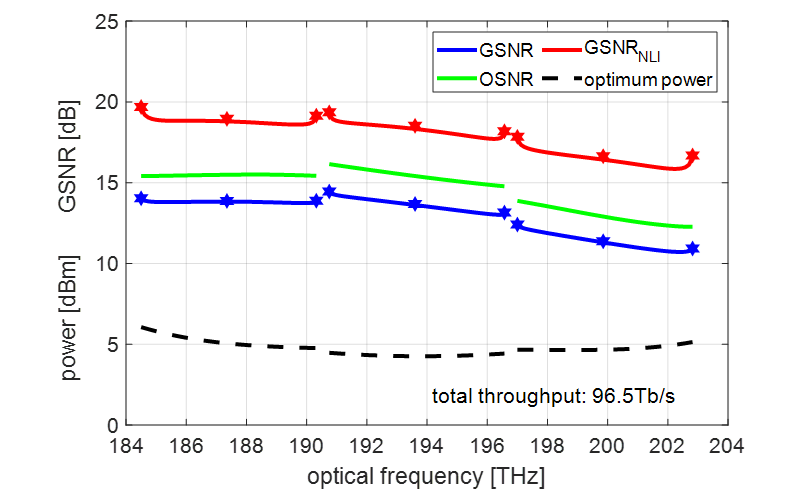}
    \caption{Same as Fig.~\ref{fig:results}(c), but with ISRS turned off.}
    \label{fig:results_ISRSoff}
\end{figure}

The cause of the somewhat disappointing performance of the S-band is its less favorable propagation conditions (higher loss, higher non-linearity coefficient, lower dispersion) and strong ISRS, transferring substantial power especially from the S-band to the L-band. To check specifically for the impact of ISRS on this system, we re-ran the optimization, turning ISRS off. The result is shown in Fig.~\ref{fig:results_ISRSoff}. Quite remarkably, the optimum launch power is now essentially flat, while GSNR has a peak-to-peak swing of only 3.5 dB. Throughput is however only 3.8\% larger than with ISRS on, confirming that the disappointing performance of the S-band in Fig.~\ref{fig:results}(c) is due more to the unfavorable propagation conditions listed above than to ISRS.
Interestingly, Fig.~\ref{fig:results_ISRSoff} shows that, in the absence of ISRS, optimizing for maximum throughput brings about the natural emergence of the so-called `3dB-rule', which states that the best system operating condition is achieved for $\rm OSNR$ being 3 dB lower than $\rm GSNR_{NLI}$ or, equivalently, ${ P_{\rm ASE}=2 P_{\rm NLI}}$ \cite{2011_Gabriella}, \cite{2011_Grellier}. This ratio is approximately found throughout the three bands. However, when ISRS is turned on, the 3dB-rule is clearly no longer optimal. Fig.\ref{fig:results}(c) shows the L-band channels operating essentially in linearity whereas the S-band ones are in deeper non-linearity than the 3dB-rule.

The disappointing performance of the S-band in Fig.~\ref{fig:results}(c) raises the issue whether some change can be enacted to improve it. It appears reasonable that  some dedicated Raman amplification could help. 
We therefore turned on three Raman pumps and optimized both their power and their frequencies. We set a max power constraint of 24 dBm for two of the pumps and 27 dBm for the third, to avoid exceeding 1 Watt of total power. We also constrained pump frequencies to stay above 211.5 THz (with no upper limit). Note that the CFM takes into account both ASE and NLI produced by backward Raman amplification, as well as pump depletion and ISRS among pumps.

\begin{figure}
    \centering
    \includegraphics[width=0.85\linewidth]{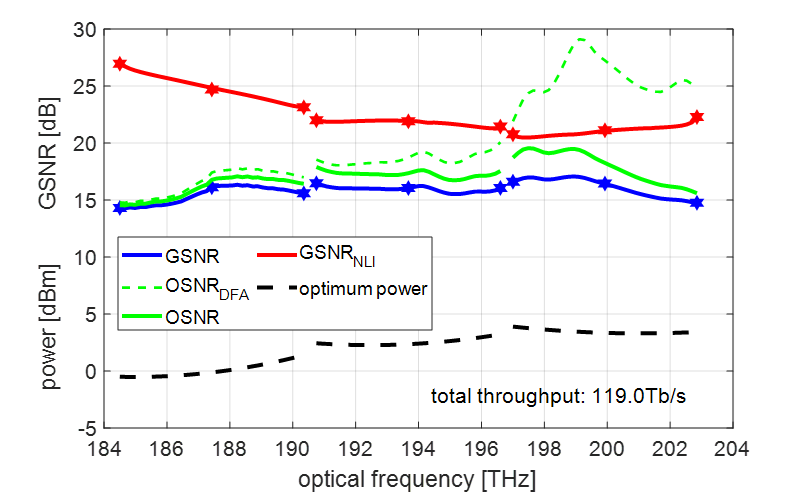}
    \caption{Same as Fig.~\ref{fig:results}(c), but with optimized backward Raman amplification for the S-band (see text for details on pumps). The dashed green curve represents $\rm OSNR$ where  ASE from lumped amplification (DFAs) only is accounted for.}
    \label{fig:results_pump}
\end{figure}

The results are shown in Fig.~\ref{fig:results_pump}. The three Raman pumps frequency and power settled at: 212.2 THz, 23.1 dBm; 213.8 THz, 23.1 dBm; 217.3 THz, 26.0 dBm.  The GSNR improves drastically in the S-band, and also substantially in the high C-band, vs.~Fig.~\ref{fig:results}(c). Its flatness also greatly improves too, to ±1.4 dB across the whole spectrum. To clarify why, Fig.~\ref{fig:results_pump} shows the quantity $\rm OSNR_{DFA}$ (dashed green line), which accounts for the ASE noise produced by DFAs \textit{only}. Comparing it with $\rm OSNR$ (solid green line), which accounts for ASE from both DFAs and Raman, it is apparent that in the S-band almost all of ASE comes from Raman amplification. Such ASE is however much less than produced by DFAs in the DFA-only set-up of Fig.~\ref{fig:results}(c). In fact, we found the average equivalent noise-figure of a fictitious DFA, producing the same gain and the same ASE as backward Raman, to be about $-1.5$~dB (at 200 THz). 

Thanks to these improvements, the overall throughput is now 119.0~Tb/s, 3.5x the result of the 6THz C-band alone. As compared to legacy 4.4 THz C-band (27Tb/s), the result is better than 4x. As a side remark, here too the optimization leads far away from the 3-dB rule, with the L-band channels being almost in linearity, while  most fo the S-band ones are in non-linear regime ($\rm GSNR_{NLI}<OSNR$).

\begin{figure}
    \centering
    \includegraphics[width=0.85\linewidth]{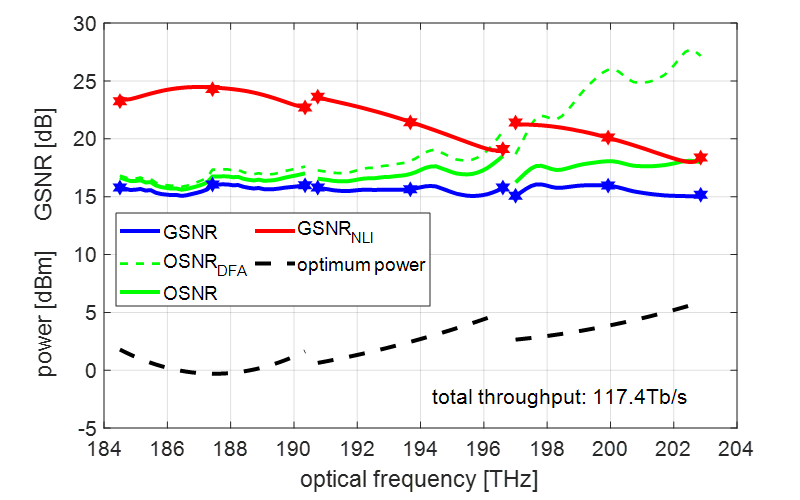}
    \caption{Same as Fig.~\ref{fig:results_pump} except the optimization objective function Eq.~(\ref{eq:objective_function_flat}) is used instead of Eq.~(\ref{eq:objective_function}), to enhance GSNR flatness.}
    \label{fig:results_pump_flatness}
\end{figure}

Finally, we decided to modify the objective function to not only seek throughput maximization, but also GSNR flattening:
\begin{equation}
    f_{\rm obj} = {\rm mean}\left( {\rm IR}_{\rm Rx}^{n} \right)-\mid {\rm IR}_{\rm Rx}^{\max}-{\rm IR}_{\rm Rx}^{\min} \mid    
    \label{eq:objective_function_flat}    
\end{equation}   
The result of the optimization is shown in Fig.~\ref{fig:results_pump_flatness}. The GSNR flatness is now remarkable $\pm 0.5$ dB across all spectrum. The three Raman pumps frequency and power settled at: 212.5 THz, 22.7 dBm; 214.8 THz, 22.9 dBm; 217.3 THz, 25.7 dBm.


\section{Conclusion}
The availability of reliable and fast CFMs allows to design and optimize possible multiband system solutions, properly accounting for the frequency dependence of all parameters, as well as for ISRS and Raman amplification. The results obtained through such optimized designs are encouraging regarding the potential benefits of the addition of the S-band in future long-haul systems. Moderate Raman amplification focused on the S-band appears to provide a very beneficial effect through all bands, allowing a potential 3x throughput increase with respect to super-C-band EDFA systems and more than 4x vs. standard (4.4-4.8 THz) EDFA C-band systems. As a side result, both with and without Raman amplification, the optimum propagating conditions of C+L+S systems are far away from the so-called `3dB rule'.

\end{document}